\documentclass[12pt]{iopart}
\usepackage{bm}
\usepackage{amssymb}
\usepackage{colordvi}
\usepackage{graphicx}
\usepackage{color}
\usepackage{hyperref}
\newcommand{\up}{\uparrow}
\newcommand{\down}{\downarrow}
\def\spin{{\mbox{\boldmath{$\hat{\sigma}$}}}}

\newcommand{\be}{\begin{equation}}
\newcommand{\ee}{\end{equation}}
\newcommand{\bea}{\begin{eqnarray}}
\newcommand{\eea}{\end{eqnarray}}

\newcommand{\vep}{\varepsilon}

\def\nn{\nonumber}

\def\grad {\mbox{\boldmath$\nabla$\unboldmath}}

\begin{document}

\title[Non-Abelian toplogical superconductors from topological semimetals]{Non-Abelian toplogical superconductors from topological semimetals and related
  systems under superconducting proximity effect}


 \author{Jian-Hua Jiang$^1$, Si Wu$^2$}
\address{$^1$ Department of Condensed Matter Physics, Weizmann Institute of
   Science, Rehovot 76100, Israel}
\address{$^2$ Department of Physics and Astronomy, University of
   Waterloo, Waterloo, Ontario N2L 3G1, Canada}

\date{\today}

\begin{abstract}
Non-Abelian toplogical superconductors are characterized by the
existence of {zero-energy} Majorana fermions bound in the quantized vortices. This is a
consequence of the nontrivial bulk topology characterized by an {\em
  odd} Chern number. It is found that in topological semimetals with a
single two-bands crossing point all the gapped superconductors are
non-Abelian ones. Such a property is generalized to related but more
generic systems which will be useful in the search of non-Abelian
superconductors and Majorana fermions.
\end{abstract}



\submitto{\JPCM}

\maketitle

\section{Introduction} Since the discovery of the quantum Hall
effects\cite{QH}, efforts devoted to understanding various topological
states of matter and their phase transitions greatly enriched the
study of condensed matter physics\cite{TKNN,Wen,TIreview}. One of the
significant aspects is that two topologically distinct states, which
can {\em not} be adiabatically connected to each other, can have the
{\em same} symmetry. This breaks down the Landau-Ginzburg paradigm of
phase transitions. {Besides} exotic excitations obeying {\em
  non-Abelian statistics} have been found in genuine and model
systems\cite{MR,Kitaev,Wen,Volovik,Maeno}. In one known {class} the
non-Abelian topological orders are closely related to the fermionic
superconducting (or superfluid) pairing states with {\em odd} Chern
numbers\cite{ReadGreen,Ady}. {Protected by the topology there is a
zero-energy Majorana fermion in each quantized vortex or on the
boundary between the system and a normal (topologically trivial) system.}
The quantum degeneracy of the ground states with $2N$ quantized
vortices (far away from each other) is $2^N$. {Winding
  between these vortices induces an unitary transformation in the $2^N$
  dimensional Hilbert space which leads to the non-Abelian
  statistics. It has been proposed that} non-Abelian excitations such
as Majorana fermions can be exploited for the topological protected
quantum computations\cite{Ivanov,Nayak,tqc,atom-qc}.

Besides the known non-Abelian topological orders in fractional quantum
Hall systems\cite{MR}, spin liquids\cite{Kitaev,Wen}, $^3$He
films\cite{Volovik}, and Sr$_2$RuO$_4$\cite{Maeno}, recently there are
theoretical proposals for non-Abelian topological orders {on} the
surface of topological insulators\cite{TI} and {in} spin-orbit
coupled two-dimensional {electron/hole systems}\cite{soc,Jason,QiQAH}
under the superconducting proximity effect as well as in ultracold
atomic gases\cite{cold-atom}. {Time-reversal symmetry breaking is
  necessary for the nonzero Chern number which
  can be realized by the magnetic field or via time-reversal symmetry
  breaking superconducting order or Zeeman type interactions. On the
  surface of a topological
insulator under $s$-wave superconducting proximity effect although
there is no time-reversal symmetry breaking, the effective ``vacuum''
of the system is an massive Dirac electron system which breaks the
time-reversal symmetry. The Chern number difference between the system
and the vacuum (as will also be shown later in this work) is $\pm 1$ which
protects the zero-energy Majorana fermion in each vortex or on the
boundary between the system and the effective vacuum\cite{detect}. In
fact to detect the Majorana fermions in such systems a Zeeman type
interaction is usually invoked to induce the effective
``vacuum'' somewhere\cite{detect}. Besides the vortices are usually induced by an external
magnetic field.} {More}  recently signatures of
Majorana fermions are observed in spin-orbit coupled one-dimensional
quantum wires proximate to superconductors\cite{1d,exp}. Inspired by
the searches for Majorana fermions\cite{review}, in this work we study
the topological properties of superconducting states in generic
semimetals and related systems under general superconducting proximity
conditions.

Topological semimetals studied here are systems consist of two-bands
crossing points (TBCPs) around the Fermi level which
can be viewed as ${\bf k}$-space vortices\cite{Volovik,Wilczek}. Away
from the TBCP the two bands do not overlap unless through other
TBCPs. In two-dimensions the TBCP has co-dimension two and
carries an integer winding number which can be
computed through the Berry phase\cite{Volovik,Sun},
\be
N_w = \frac{1}{\pi}\oint_{{\cal
    C}} d{\bf k}\cdot \langle \Psi({\bf k})|i\grad_{\bf k}|\Psi({\bf
  k})\rangle .
\ee
Here ${\cal C}$ is an anti-clockwise path enclosing the
TBCP and $\Psi({\bf k})$ is the wavefunction (single-valued and
continuous) in the band with energy above (or below) the
TBCP. Concrete examples are Dirac cones and quadratic band
crossings\cite{Sun} [see Fig.~1 (a) and (b)] where the winding number
is $N_w=\pm 1$ and $\pm 2$ respectively. {Here the integer
  $N_w$ is only defined for the band crossing (whenever the band
  crossing is gapped $N_w$ will no longer be an
  integer). The winding number $N_w$ characterizes the
  TBCP\cite{Volovik,Wilczek} and will be used to classify different
  situations in this study.} The discussions hereafter
will be split into two cases: (A) when $N_w$ is even, and (B) when $N_w$ is
odd. Time-reversal ${\cal T}$ symmetry is
imposed for both cases. For concreteness case A is restricted to
systems with zero angular momentum where the time-reversal operator is
${\cal T}={\cal K}$ (${\cal K}$ is complex conjugation), whereas
case B for spin-half systems where ${\cal T}={\cal K} i\sigma_y$.

The findings in this work are: (i) When there is a single TBCP,
for both cases A and B, {\em all} gapped superconducting states
are {\em non-Abelian}. Namely, the bulk Chern number is {\em
  odd}. (ii) The same conclusion holds when the TBCP are gapped due
to time-reversal symmetry breaking but inversion symmetric
perturbations. (iii) It also holds when the TBCP are gapped and
deformed (such that the two bands eventually evolves in the same
direction in energy) given that only one band crossing the Fermi
level. (iv) The discussion is further extended to situations where the
time-reversal symmetry is broken and those where there are multiple
such TBCPs.

The paper is organized as follows: In Sec.~II we discuss the situation
with a single TBCP for cases A and B. In Sec.~III the situations
when the TBCP is gapped (and deformed) are studied. In Sec.~IV we
develop more generalizations. We conclude in Sec.~V. All the
discussion are restricted in the weak pairing regime {which is
relevant to proximity induced pairing orders}.

\section{Topological semimetals with a single TBCP}

\subsection{Case A}
In spinless (or spin-polarized) many-fermion systems in 2D lattices
with multiple orbits in an unit cell {with inversion symmetry}, there
can be TBCPs with even winding numbers. Around such a TBCP the
Hamiltonian can be generally written as
\be
H_0({\bf k}) = h_0({\bf k})\hat{\sigma}_0 + h_x({\bf k}) \hat{\sigma}_x
+ h_z({\bf k})\hat{\sigma}_z .
\ee
Here the Pauli matrices, $\sigma_x$ and $\sigma_z$, act on the Wannier
orbits (pseudo-spins), and $\sigma_0$ is the $2\times 2$ identity
matrix. Due to time-reversal symmetry, the TBCP can only be at a
time-reversal invariant momentum ${\bf K}$ when there is only a single
such TBCP. ${\bf k}$ is the wavevector measured from ${\bf
  K}$. $h_{\nu}(-{\bf k})=h_{\nu}({\bf k})$ for $\nu=0,x,z$ and
$h_y({\bf k})\equiv 0$ due to time-reversal {and inversion}
symmetry. The spectrum is $\vep_{{\bf  k}\pm} = h_0({\bf k}) \pm \sqrt{h_x^2({\bf k}) +
  h_z^2({\bf k})}$. For semimetals, $|h_0({\bf k})|<\sqrt{h_x^2({\bf
    k}) + h_z^2({\bf k})}$ and $h_\nu=0$ at $k=|{\bf k}|=0$. The
eigenstates of $H_0({\bf k})$ are
\bea
|u_{+}({\bf k})\rangle &=&
\frac{1}{2}[(e^{-i\phi_{\bf k}} + 1)|\up\rangle + i(e^{-i\phi_{\bf k}}
- 1)|\down\rangle],\nn\\ |u_{-}({\bf k})\rangle &=&
\frac{1}{2}[i(e^{-i\phi_{\bf k}} - 1)|\up\rangle -(e^{-i\phi_{\bf k}}
+ 1)|\down\rangle,
\eea
with $\phi_{\bf k} = {\rm Arg}[h_z({\bf k})+ih_x({\bf k})]$. The
winding number of the TBCP {is calculated through Eq.~(1)} as
\be
N_w = \frac{1}{2\pi} \oint_{{\cal C}} d\phi_{\bf k} .
\ee
This has a transparent physical meaning: $N_w2\pi$ is the winding
angle of ${\bf h}$ and that of the pseudo-spin direction. The winding number
can only be an even integer as $h_{\nu}(-{\bf k})=h_{\nu}({\bf k})$.

\begin{figure}[htb]
\includegraphics[height=2.8cm]{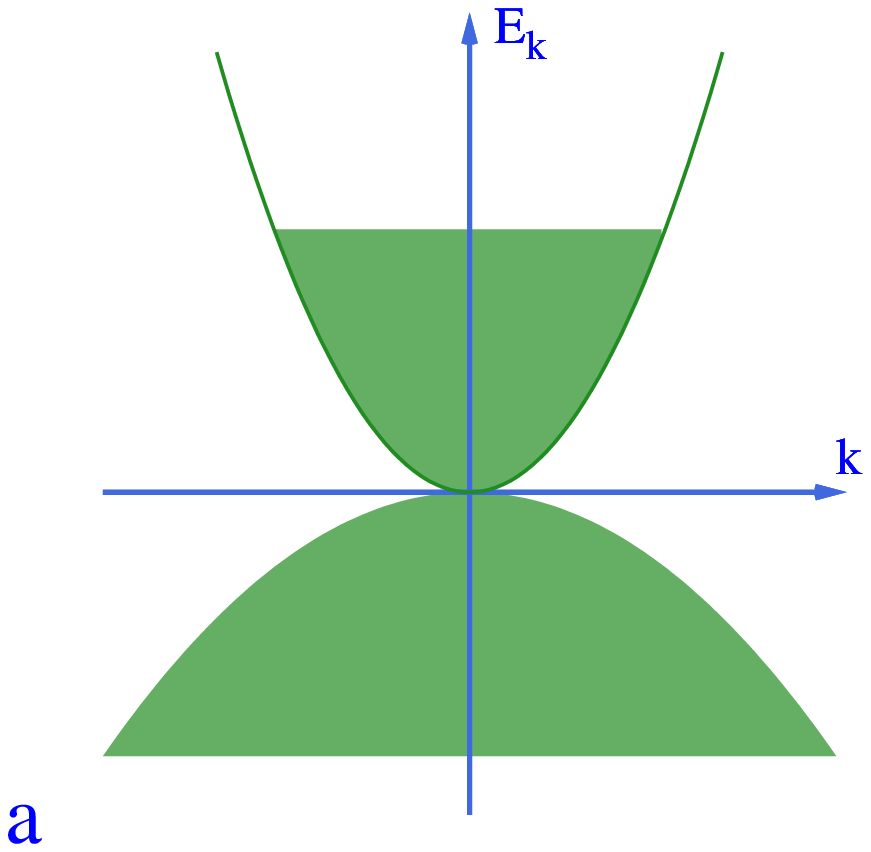}\includegraphics[height=2.8cm]{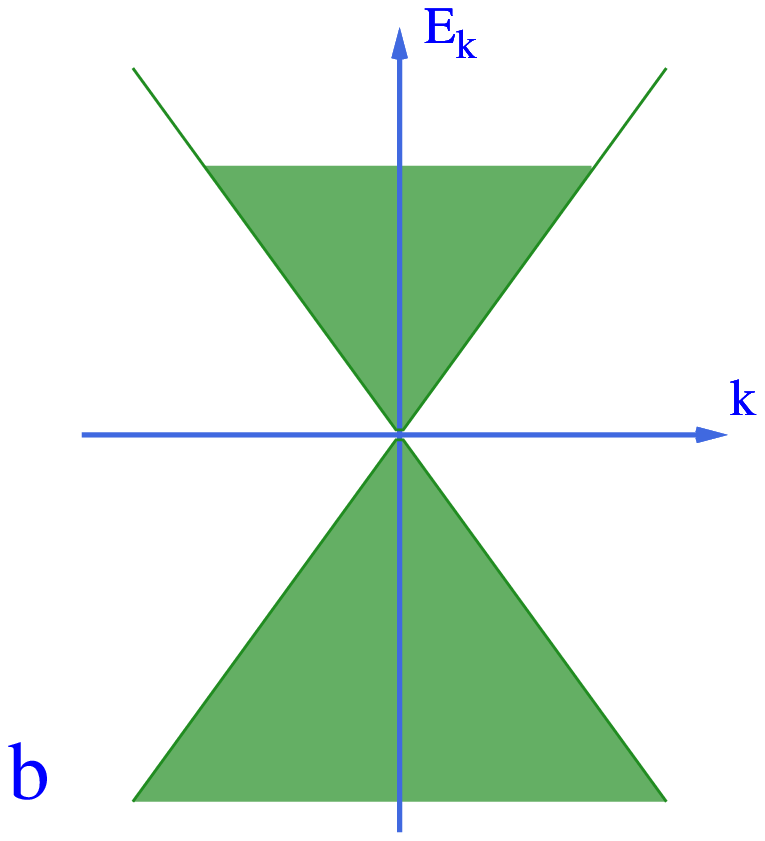}
\includegraphics[height=2.8cm]{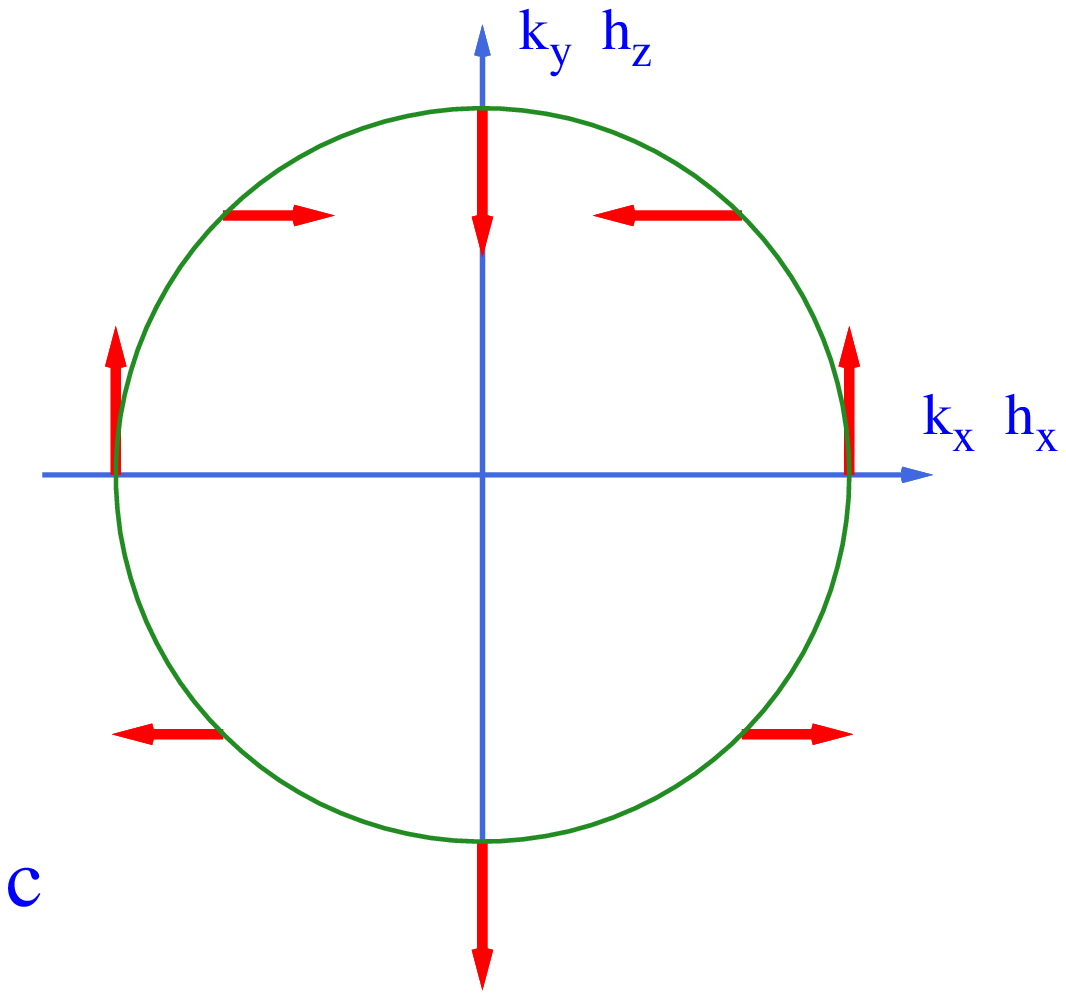}\includegraphics[height=2.8cm]{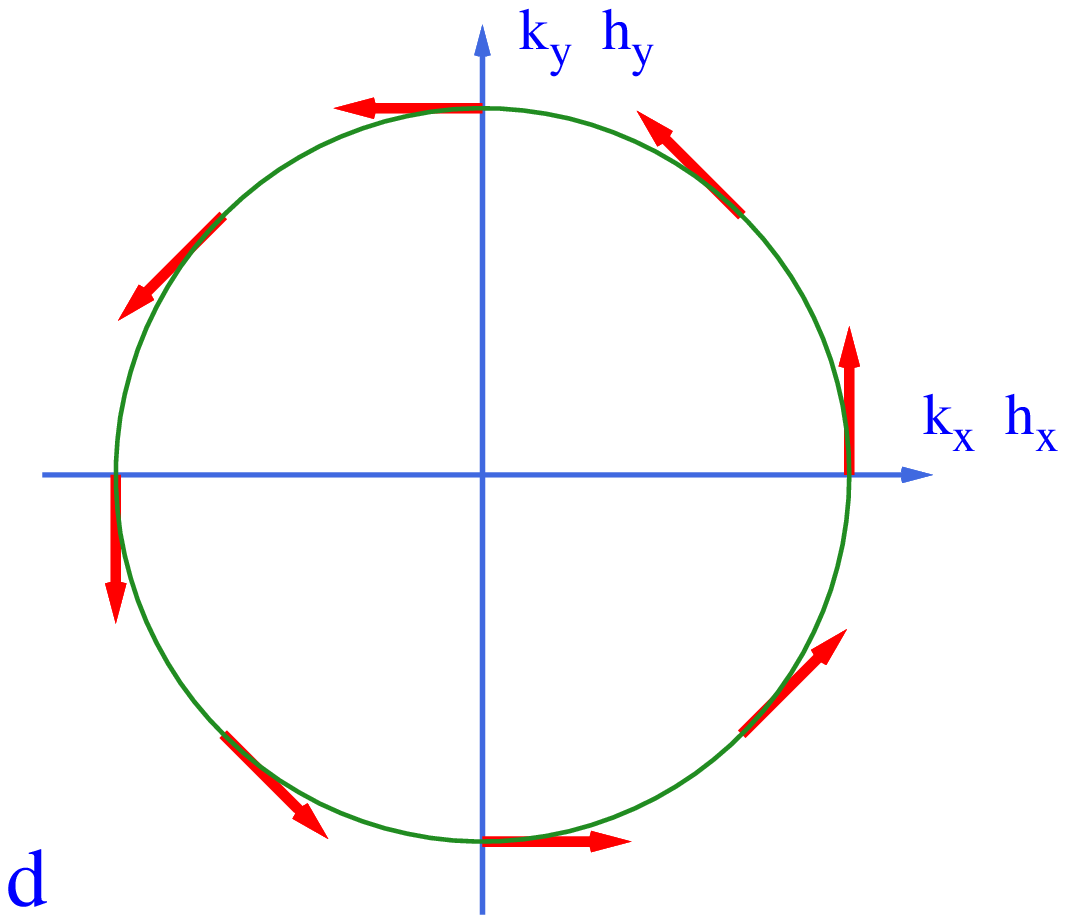}
\caption{(Color online) Illustration of the quadratic band crossing
  (a) and the Dirac cone (b) with Fermi surface above the TBCP. Dark
  region denotes the filling of electrons. Illustration of the
  direction of the field $(h_x,h_z)$ or $(h_x,h_y)$ (also represents
  the pseudo-spin or spin  direction) at Fermi surface for
  the quadratic band crossing (c) and the Dirac cone (d).}
\label{fig1}
\end{figure}

One example of such TBCP systems is the quadratic band crossing in the
checkerboard lattices\cite{Sun,sun2,our}, where in the vicinity of
${\bf K}=(\pi,\pi)$, $h_0({\bf k}) = t_0k^2$, $h_x({\bf k}) =
2t_xk_xk_y$, $h_z({\bf k}) = t_z(k_x^2-k_y^2)$, and $h_y({\bf
  k})\equiv 0$ with $t_0$, $t_x$, and $t_z$ being the band
parameters. The system is a semimetal with winding number $N_w=2{\rm
  sgn}(t_xt_z)=\pm 2$ when $|t_0|<|t_x|, |t_z|$.

The general form of the Bogoliubov-de Gennes (BdG) Hamiltonian for the
system is $H=\frac{1}{2}\sum_{\bf k}\Psi^\dagger({\bf k})\mathcal{H}_{{\bf
    k}}\Psi({\bf k})$ with $\Psi({\bf k})=\left(\psi_\up({\bf
  k}),\psi_\down({\bf k}),\psi_\up^{\dagger T}(-{\bf
  k}),\psi_\down^{\dagger T}(-{\bf k})\right)^T$ and
\be
{\cal H}_{\bf k} = \left[\begin{array}{ccc}
    H_0({\bf k}) -\mu & &
    -\hat{\Delta}({\bf k}) \\
       \hat{\Delta}^\ast(-{\bf k}) & & -H_0^\ast(-{\bf k}) + \mu
     \end{array} \right] .
   \label{mf}
\ee
Here $\hat{\Delta}({\bf k}) = i\Delta_0({\bf k})\hat{\sigma}_y +
\Delta_z({\bf k})\hat{\sigma}_x + i\Delta_y({\bf k})\hat{\sigma}_0 -
\Delta_x({\bf k})\hat{\sigma}_z$ is the general form of
superconducting pairing interaction when the Cooper pair have
zero-angular momentum. $\Delta_0$ and $\Delta_\nu$ ($\nu=x,y,z$)
represent the singlet and triplet pairings respectively.

In the weak pairing regime, $|\Delta_\nu|\ll |\mu|$, only the pairing
interaction between nearly degenerate states are important, whereas
that between states far away can be ignored. The pairing properties
can then be studied by projecting the original Hamiltonian
into the subspace spanned by the band that crosses the Fermi level. To
the leading order, the projected BdG Hamiltonian is $H_{\rm
  PBdG}=\frac{1}{2}\sum_{\bf k}\Psi_P^\dagger({\bf
  k})\mathcal{H}_{{\bf  k}}^{P} \Psi_P({\bf k})$ with $\Psi_P({\bf
  k})=(c_{{\bf k}\pm}, c_{-{\bf  k}\pm}^{\dagger})^T$ and
\be
{\cal H}_{\bf k}^{P} = \left[ \begin{array}{ccccccccc}
    \vep_{{\bf k}\pm} - \mu & & \Delta_{\rm eff}({\bf k}) \\
    \Delta_{\rm eff}^{\ast}({\bf k}) &  & -\vep_{{\bf k}\pm} + \mu  \\
\end{array} \right] .
\label{pbdg}
\ee
Here the $+$ and $-$ indices are for the $\mu>0$ and $\mu<0$ cases
respectively and
\be
\Delta_{\rm eff}({\bf k}) = e^{i\phi_{\bf k}}\left[i\Delta_y - \frac{1}{2}{\rm sgn}(\mu)\sum_{\pm}
(\Delta_x \pm  i\Delta_z) e^{\pm i\phi_{\bf k}}\right] .
\label{Dqbt}
\ee
The eigenstates can then be obtained by directly diagonalizing the
above Hamiltonian. The Chern number is given by\cite{rmp}
\be
N_C = \sum_n \frac{1}{2\pi}\int
d{\bf k}~ {\bf e}_z\cdot [\grad_{\bf k}\times
  \langle\Psi_{n}|i\grad_{\bf k}|\Psi_{n}\rangle] ,
\label{N_C}
\ee
where $\Psi_{n}$ are the wavefunctions of the occupied bands.
Direct calculation yields [for details, see Appendix]
\be
N_C = \left. {\rm sgn}(\mu) \int_0^{2\pi}\frac{d{\theta_{\bf k}}}{2\pi}
\partial_{\theta_{\bf k}}\theta_{\Delta}({\bf k})\right |_{\rm FS} ,
\label{nc}
\ee
where $\theta_{\Delta}({\bf k})={\rm Arg}[e^{-i\phi_{\bf k}}\Delta_{\rm eff}({\bf
  k})]$. That is, the Chern number is nothing but the winding number
of $e^{-i\phi_{\bf k}}\Delta_{\rm eff}({\bf k})$ at the Fermi surface
(denoted as 'FS' above). Physically this is due to the fact that the
superconducting gap is only opened at the Fermi surface in the weak
pairing regime\cite{QiZ2}. From Eqs.~(\ref{Dqbt}) and (\ref{nc}),
the effect of the Fermi surface Berry phase on the Chern number
is clearly visible.

The error of the eigenstates obtained from the projected Hamiltonian
is on the order of ${\cal O}(|D|/|\mu|)$. However, this induces
{\em no} error in the calculated Chern number due to its topological
nature. Namely one always can adiabatically tune the pairing
interaction $\Delta_\nu\to \alpha \Delta_\nu$, via one scaling factor
$\alpha$, to sufficiently small to reduce the error without closing
the superconducting gap. As the gap is not closed, the Chern number
does not change. Hence the error of the calculated Chern number can be
infinitesimally small when $\alpha\to 0$. Note that Eq.~(\ref{nc})
does not depend on $\alpha$. Therefore, there is no error in the Chern
number calculated via the projected BdG Hamiltonian in the weak
pairing regime.

A crucial observation is that the winding number of
$e^{-i\phi_{\bf k}}\Delta_{\rm eff}({\bf k})$ at the Fermi surface can
only be {\em odd} when it is well-defined. This is because the winding
number of $\Delta_\nu$ ($\nu=x,y,z$) is {\em odd} while that of
$e^{\pm i\phi_{\bf k}}$ are {\em even}. Hence the Chern number $N_C$ can
only be {\em odd}. Therefore {\em all} the gapped superconducting
states in case A are {\em non-Abelian} ones.

\subsection{Case B}
TBCPs with an odd winding number, such as Dirac cones, can appear in spin
half fermionic systems\cite{TIreview}. The fermion doubling theorem
states that in 2D lattice systems there can only be an {\em even}
number of such TBCPs\cite{Nilsen}. However at the surface of strong
topological insulators there can be an odd number of such TBCPs. The
concerned systems have a single such TBCP at a time-reversal invariant
momentum ${\bf K}$ due to time-reversal symmetry. Rather than
$N_w=\pm 1$ for Dirac cone, $N_w$ can be any {\em odd} integer here.
The general Hamiltonian around such a TBCP is
\be
H_0({\bf k}) = h_0({\bf k})\sigma_0 + {\bf h}({\bf k})\cdot\spin ,
\label{odd-g}
\ee
where the Pauli matrices now denote true-spin and ${\bf k}$ measured
from ${\bf K}$. $|h_0|<|{\bf h}|$ and $|{\bf h}|=0$ at ${\bf k}=0$ [we also
set $h_0({\bf k}=0)=0$] so that the system is a semimetal. We choose the
coordinates so that $h_z({\bf k})=0$, i.e., the winding axis is along the
$z$-direction. The spectrum is $\vep_{\pm{\bf k}}=h_0({\bf k})\pm
\sqrt{h_x^2+h_y^2}$ and the eigenstates are
$|u_{\pm}(\mathbf{k})\rangle = \frac{1}{\sqrt{2}} (e^{-i\psi_{\bf
    k}}|\up\rangle \pm |\down\rangle)$ with $\psi_{\bf k}={\rm
  Arg}[h_x({\bf k})+ih_y({\bf  k})]$. {Using Eq.~(1)} one finds
\be
N_w = \frac{1}{2\pi} \oint_{{\cal C}} d\psi_{\bf k}  .
\ee
The winding number can only be an {\em odd} integer as ${\bf h}(-{\bf
  k})=-{\bf h}({\bf k})$ according to time-reversal symmetry.

Following the argument in previous section, in the weak pairing regime
one can study the topological property of the system via the
projected BdG Hamiltonian [Eq.~(\ref{pbdg})]. Here
\be
\Delta_{\rm eff}({\bf k}) = e^{i\psi_{\bf k}}\Big[{\rm sgn}(\mu)\Delta_0 +
\frac{1}{2}\sum_{\pm} (\Delta_x \mp i\Delta_y ) e^{\pm i\psi_{\bf k}}\Big] .
\ee
The Chern number $N_C$ is the winding number of $e^{-i\psi_{\bf
    k}}\Delta_{\rm eff}({\bf k})$ at the Fermi surface as in
Eq.~(\ref{nc}). It is noted that the winding number of $e^{-i\psi_{\bf
    k}}\Delta_{\rm eff}({\bf k})$ can only be {\em even}.
Therefore the Chern number can only be {\em even}.

As the concerned system lives only on the boundary of two
three-dimensional systems with distinct $Z_2$ topology, it does not have
well-defined edges\cite{TIreview}. One way to circumvent this problem
is to circulate the superconducting state with a ferromagnetic
insulating state with the same $H_0({\bf k})$ but with a
magnetization along $z$-direction $M\sigma_z$\cite{TI}. When
$|M|>|\mu|$\cite{note1}, the quasi-particles can not propagate into
the ferromagnetic region. On the boundary between the superconducting
region and the ferromagnetic one, there are gapless Majorana edge
states. The ferromagnetic insulating state is topologically equivalent
to a superconducting massive Dirac fermion system with $|M|>|\mu|$. It
has a Chern number of ${\rm sgn}(M)N_w$. For instance, there may be
$n_c$ clockwise moving edge states and $n_a$ anti-clockwise moving
edge states. According to bulk-edge correspondence, $n_c-n_a=N_C-{\rm
  sgn}(M)N_w$. The difference $n_c-n_a$ is fixed by topology and is
always {\em odd} as $N_C-{\rm sgn}(M)N_w$ is odd. Therefore the total
number of edge states ${\cal N}_{\rm edge}=n_c+n_a$ is definitely {\em odd}.

The above analysis can also be applied to the Majorana bound states in
the core of a quantized vortex, which can be viewed as edge states
live in the small circular edge of the vortex with vacuum at the
center\cite{ReadGreen}. As the boundary condition at the center
does not affect the existence of the zero-energy Majorana bound state,
it can be tuned that the vacuum at the center is a superconducting
massive Dirac fermion with $|M|>|\mu|$. Therefore there are
${\cal N}_{\rm edge}$ number of Majorana states in core of a quantized
vortex. In reality, there are inevitable mixing between those states
(e.g., due to disorder) and interactions between the Majorana
fermions, which lift the degeneracy. However, the particle-hole
symmetry guarantees the existence of {\em one} zero-energy Majorana
bound state when ${\cal N}_{\rm edge}$ is odd. Accordingly, {\em all}
the gapped superconducting pairing states here are {\em non-Abelian}
ones since ${\cal N}_{\rm edge}$ is definitely {\em odd}. This
argument (essentially the same as that in
Ref.~\cite{ReadGreen}), verifies the existence of the Majorana
zero modes in the vortex core {\em without explicitly solving the
  Schr\"odinger equation for the quasi-particle spectrum in a vortex}
as such property is essentially dictated by the bulk
topology\cite{ReadGreen}. Recent theories\cite{majorana} also present
additional proofs on such relation of the number of Majorana zero
modes to the Chern numbers.

\begin{figure}[htb]
\includegraphics[height=2.7cm]{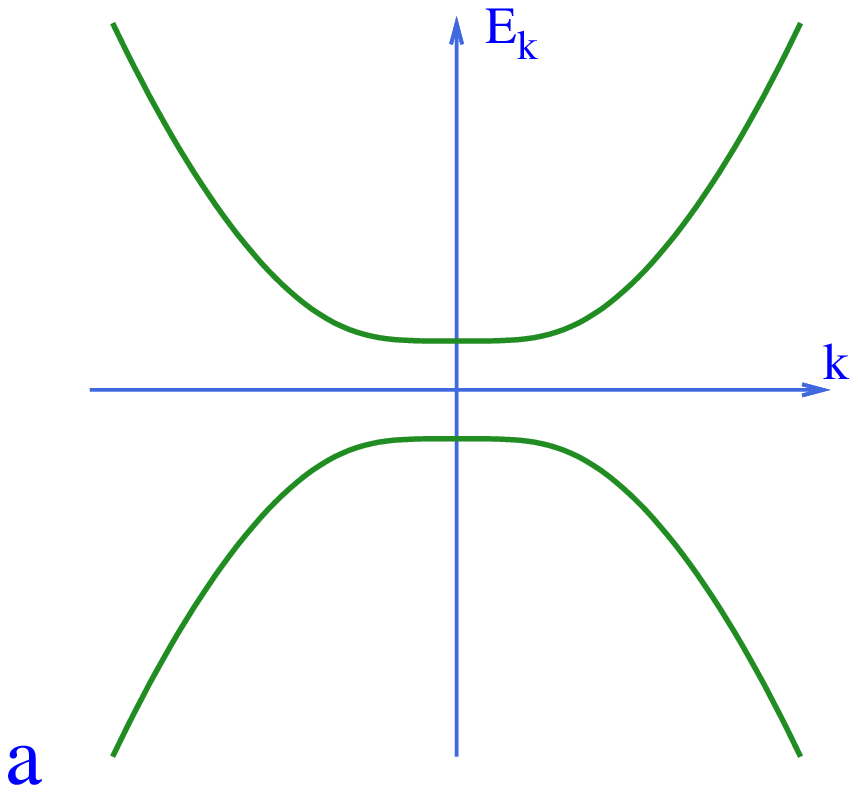}\includegraphics[height=2.7cm]{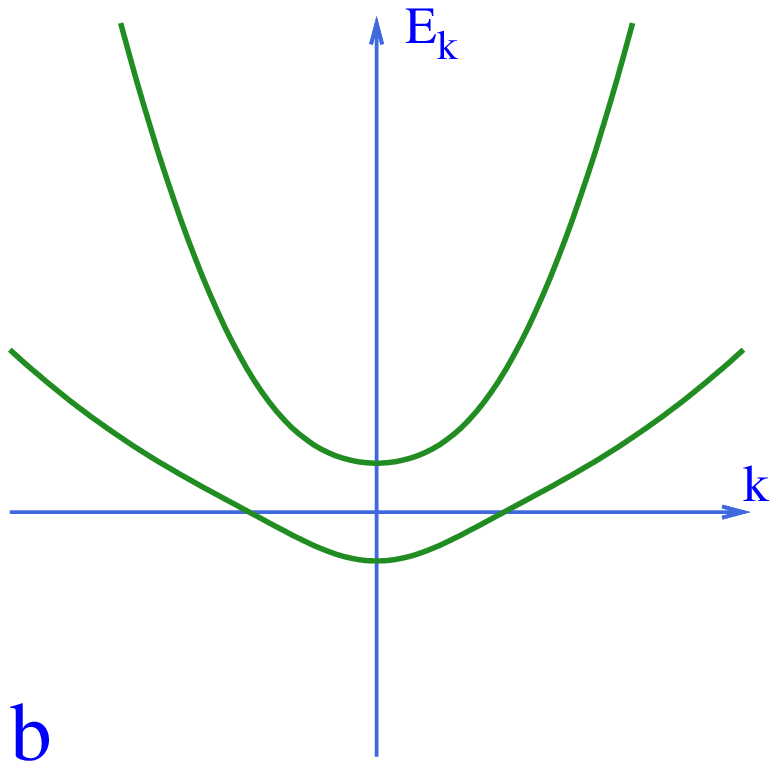}
\includegraphics[height=2.7cm]{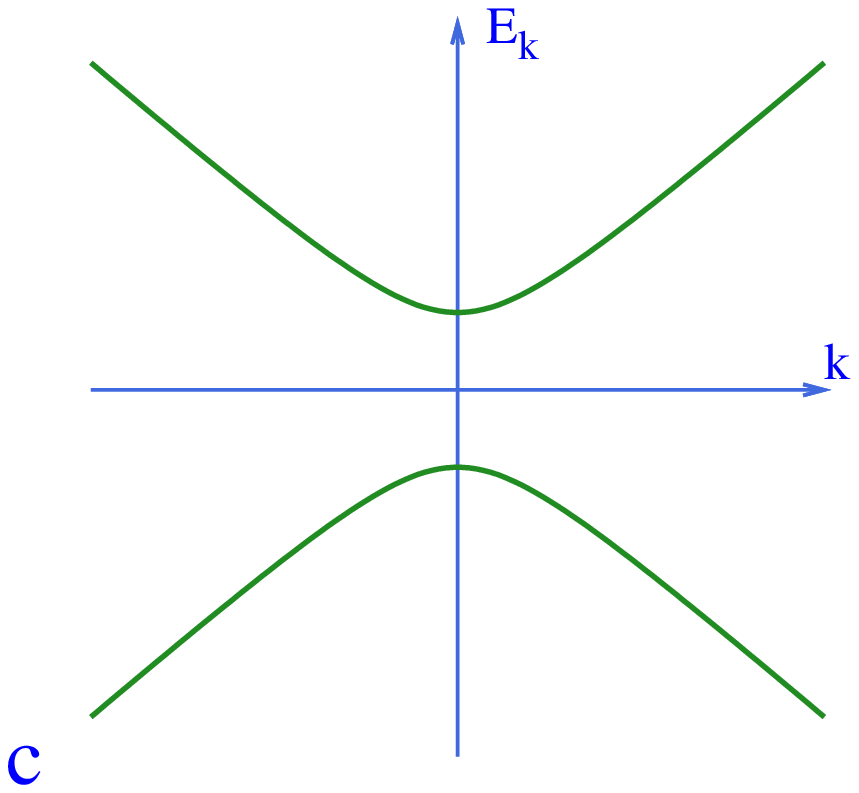}\includegraphics[height=2.7cm]{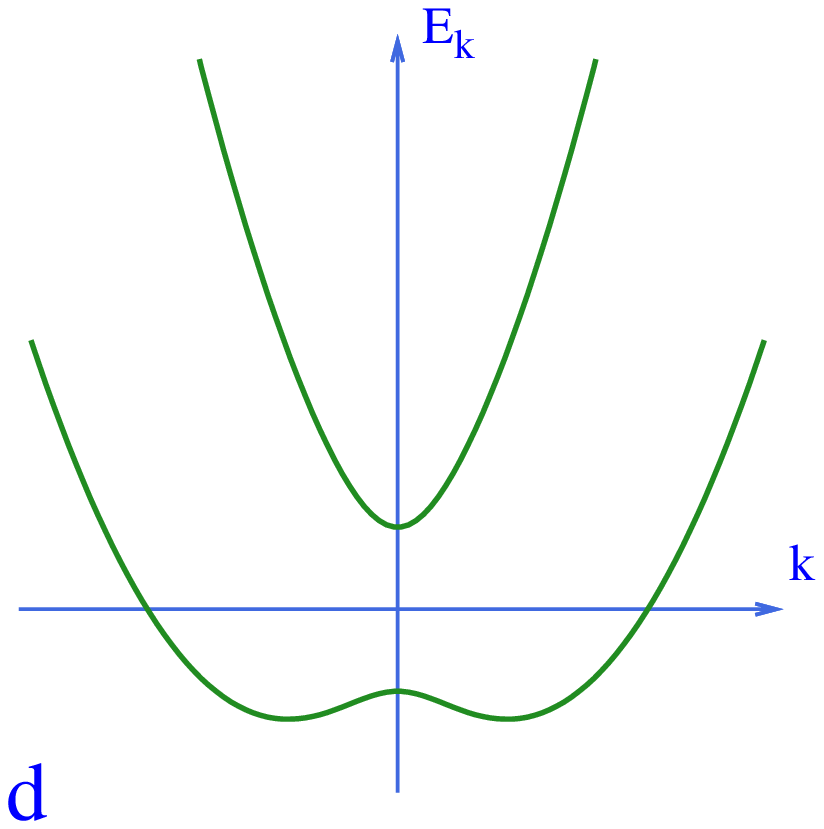}
\caption{(Color online) Illustration of the dispersions in (a) gapped
  (b) gapped and deformed  quadratic band crossing systems as well as (c)
  gapped and (d) gapped and deformed Dirac cone systems.}
\label{fig2}
\end{figure}

\section{Systems with a single gapped/deformed TBCP}
The general Hamiltonian is given by Eq.~(\ref{odd-g}) with all the
$h_\nu({\bf k})$ ($\nu=0,x,y,z$) being nonzero. The spectrum is
$\vep_{{\bf  k}\pm} = h_0({\bf k}) \pm |{\bf h}|$ with $|{\bf h}| =
\sqrt{h_x^2({\bf k}) + h_y^2({\bf k}) + h_z^2({\bf k})}$. For case A the
eigenstates are
\bea
|u_{+}({\bf k})\rangle = \frac{1}{\sqrt{2}}\left(\begin{array}{cc}
\cos\frac{\eta_{\bf k}}{2}e^{-i\phi_{\bf k}} + \sin\frac{\eta_{\bf k}}{2} \\
i\cos\frac{\eta_{\bf k}}{2}e^{-i\phi_{\bf k}} - i
\sin\frac{\eta_{\bf k}}{2} \end{array} \right) ,\nn \\
|u_{-}({\bf k})\rangle = \frac{1}{\sqrt{2}}\left(\begin{array}{cc}
i\sin\frac{\eta_{\bf k}}{2}e^{-i\phi_{\bf k}} - i\cos\frac{\eta_{\bf k}}{2} \\
- \sin\frac{\eta_{\bf k}}{2}e^{-i\phi_{\bf k}} -
\cos\frac{\eta_{\bf k}}{2} \end{array} \right) .
\eea
where $\eta_{\bf k} = {\rm Arg}[h_y+i\sqrt{h_x^2+h_z^2}]$ and
$\phi_{\bf k}={\rm Arg}[h_z+ih_x]$. At $k=0$, $h_x=h_z=0$ whereas
$h_y\ne 0$. At large $k$, $|h_y|\ll \sqrt{h_x^2+h_z^2}$ and $\eta_{\bf
  k}\to \pi/2$. We restrict the discussion to the situations where
$h_y(-{\bf k})=h_y({\bf k})$. Hence $\eta_{\bf k}$ is an even function
of ${\bf k}$. For case B the eigenstates are $|u_{+}({\bf k})\rangle =
\cos\frac{\zeta_{\bf k}}{2}e^{-i\psi_{\bf k}}|\up\rangle +
\sin\frac{\zeta_{\bf k}}{2}|\down\rangle$ and $|u_{-}({\bf k})\rangle =
\sin\frac{\zeta_{\bf k}}{2}e^{-i\psi_{\bf k}}|\up\rangle -
\cos\frac{\zeta_{\bf k}}{2}|\down\rangle$,
where $\zeta_{\bf k} = {\rm Arg}[h_z+i\sqrt{h_x^2+h_y^2}]$ and
$\psi_{\bf k}={\rm Arg}[h_x+ih_y]$. Now at $k=0$, $h_x=h_y=0$ and
$h_z\ne 0$, whereas at large $k$, $|h_z|\ll \sqrt{h_x^2+h_y^2}$ and
$\zeta_{\bf k}\to \pi/2$. We also assume $h_z(-{\bf k})=h_z({\bf k})$
so that $\zeta_{\bf k}$ is an even function of ${\bf k}$.

Let's first consider case A. When $|h_y(0)|<|\mu|$ and $|h_0|<|{\bf
  h}|$ (i.e. only one band crosses the Fermi level) [e.g., see
Fig.~2(a)], exerting the previous technique to obtain the projected
BdG Hamiltonian (\ref{pbdg}), one finds
\bea
&&\hspace{-0.3cm} \Delta_{\rm eff}^{\pm}({\bf k}) = e^{i\phi_{\bf
    k}}\Big\{  \Delta_z [\sin\phi_{\bf k} \mp i\cos\phi_{\bf k}\cos\eta_{\bf
  k}] \pm i\Delta_y\sin\eta_{\bf k} \nn \\ &&\hspace{0.7cm}
-\Delta_x
[\cos\phi_{\bf k}\pm i\sin\phi_{\bf k}\cos\eta_{\bf k}]
\Big\} ,
\label{pair2}
\eea
where $+$ and $-$ are for the higher and lower bands respectively.
Direct calculation yields that the Chern number is still given by
Eq.~(\ref{nc}). And the property that all the gapped superconductors
are non-Abelian ones still holds since the winding numbers of $e^{\pm
  i\phi_{\bf k}}$ are always even and $\eta_{\bf k}$ is an even
function of ${\bf k}$. This is consistent with the picture that
opening a gap below or above Fermi level does not affect the
topological properties. A nontrivial situation is when the TBCP is both
gapped and deformed so that $|h_0|>|{\bf h}|$ at large $k$. In this
situation the two bands evolve in the same direction at large $k$
[e.g., see Fig.~2(b)]. When $|\mu|<|h_y(0)|$, only the lower band
crosses the Fermi level and the Chern number is
\be
N_C = \left. \int_0^{2\pi}\frac{d{\theta_{\bf k}}}{2\pi}
\partial_{\theta_{\bf k}}\theta_{\Delta}^{-}({\bf k})\right |_{\rm FS} +
{\rm sgn}[h_y(0)]N_w .\label{nc2}
\ee
Here $\theta_{\Delta}^{\pm}={\rm Arg}[e^{-i\phi_{\bf k}}\Delta_{\rm
  eff}^{\pm}({\bf k})]$. It is seen that as $N_w$ is {\em even}, the
Chern number is again {\em odd} for all the gapped states. Therefore
the system still has the nontrivial property that {\em all} the gapped
superconducting pairing states are non-Abelian ones. When the Fermi
level is such high that both bands cross it, the total Chern number is
\be
N_C = \sum_{\pm} \left. \int_0^{2\pi}\frac{d{\theta_{\bf k}}}{2\pi}
\partial_{\theta_{\bf k}}\theta_{\Delta}^{\pm}({\bf k})\right |_{\rm
FS} .
\label{even}
\ee
Hence the total Chern number becomes {\em even} (trivial or Abelian
topological superconductors) when the two bands cross the Fermi level.

Now we turn to case B. When the TBCP is gapped the effective
superconducting pairing in the two bands are
\bea
\Delta_{\rm eff}^{\pm} ({\bf k}) &=& e^{i\psi_{\bf k}}\Big[
\cos^2(\frac{\zeta_{\bf k}}{2})(\Delta_x \pm  i\Delta_y ) e^{\mp i\psi_{\bf
    k}}  \mp \Delta_0 \sin(\zeta_{\bf k})  \nn \\ &&\mbox{} +
\sin^2(\frac{\zeta_{\bf k}}{2})(\Delta_x \mp i\Delta_y ) e^{\pm i\psi_{\bf
    k}}
\Big] .
\eea
Here $+$ and $-$ again for the higher and lower bands respectively.
Again the Chern number is the same as that at $h_z=0$ when
$|\mu|>|h_z(0)|$ and $|h_0({\bf k})|<|{\bf h}|$ [see Fig.~2(c)]. When
the TBCP is gapped and deformed, so that $|h_0({\bf
  k})|>|{\bf h}|$ at large $k$ [e.g., see Fig.~2(d)]. Such a system
{\em can} exist as a two-dimensional lattice system without violating
the fermion doubling theorem. Examples in reality are the spin-orbit
coupled two-dimensional electron (hole) systems under a Zeeman (or
exchange) field $h_z$. When $|\mu|<|h_z(0)|$, only the lower band
crosses the Fermi level, one finds that
\be
N_C = \left. \int_0^{2\pi}\frac{d{\theta_{\bf k}}}{2\pi}
\partial_{\theta_{\bf k}}\theta_{\Delta}^{-}({\bf k})\right |_{\rm FS} +
{\rm sgn}[h_z(0)]N_w
\ee
with $\theta_{\Delta}^{\pm}={\rm Arg}[e^{-i\psi_{\bf k}}\Delta_{\rm
  eff}^{\pm}({\bf k})]$. Note that the winding number of
$e^{-i\psi_{\bf k}}\Delta_{\rm eff}({\bf k})$ can only be {\em even}
as the winding number of $e^{\pm i\psi_{\bf k}}$ are always {\em odd}
and $\zeta_{\bf k}$ is an even function of ${\bf k}$. Therefore the
Chern number can only be {\em odd}. When the Fermi level is higher so
that both bands cross it, the total Chern number is given by
Eq.~(\ref{even}), which is always {\em even}.

\section{More generalizations}
In this section we explore further generalizations of the results
obtained. The first generalization is that for systems with multiple
TBCPs (no matter whether they are gapped or deformed). Whenever the
superconducting pairing interaction is within each TBCP and there are
an {\em odd} number of bands crossing the Fermi level, the property that all
the gapped superconductors are non-Abelian ones should also hold. This
is because the total Chern number is the summation of the
contribution from each TBCP. Such situations can appear when every
TBCP is located at a time-reversal invariant momentum.

There is a possibility that when the time-reversal symmetry is broken
yet the inversion symmetry is not the spin-half system can have a
single TBCP with an {\em even} winding number. A general Hamiltonian
for such systems near the TBCP in the form of (\ref{odd-g}) is
\bea
&& h_0({\bf k}) = M_0 k^2, \quad h_z({\bf k}) = M_y - \beta k^2 , \nn\\
&& h_x({\bf k}) = \gamma (k_x^2 - k_y^2),\quad h_y({\bf k}) = 2\delta k_x k_y ,\label{iop}
\eea
where $M_0$, $M_y$, $\beta$, $\gamma$, and $\delta$ are band
parameters. The TBCP exists when $|h_0|<|{\bf h}|$ with $h_z\equiv 0$. It
is gapped when $M_y\ne 0$. Gapped and deformed when $M_y\ne 0$ and
$|h_0|>|{\bf h}|$ at large $k$. This situation is essentially the same
as case A. One can easily find that the Chern number can only be {\em
  odd} in such systems when there is only one band crosses the Fermi
level. This results can be further generalized to systems with
multiple such TBCPs.

\section{Candidate physical systems}
Beside the systems already found in the literature in the search of
Majorana fermions, such as systems with single Dirac cone and
semiconductor quantum wells with Rashba spin-orbit coupling, there are
many unexplored candidate systems which fit into the above
discussions. Below we list some candidates which have not yet
attracted researchers' attention.

\begin{itemize}

\item {\sl Semiconductor nanostructures with Zeeman (or exchange)
    splitting and arbitrary spin-orbit coupling.} This is essentially
  case A with a gapped and deformed TBCP. Given that the winding
  number $N_w$ of the TBCP is {\em odd} the system
  supports Majorana fermions in the vortex. This is a direct
  generalization of the studies in the literature\cite{soc}.
  Specific examples are: (i) Two-dimensional electron system with both
  Rashba and Dresselhaus spin-orbit couplings. For III-V semiconductor
  quantum wells with growth direction [001] when Rashba (Dresselhaus)
  spin-orbit coupling is dominant $N_w=1$ ($N_w=-1$). (ii)
  Two-dimensional heavy hole system where the cubic spin-orbit
  coupling leads to $N_w=\pm 3$.\cite{f-wave} When there are multiple
  sub-bands crossing the Fermi level. The total winding number is the
  summation of the winding number of each sub-band. If the total
  winding number is {\em odd} then all the gapped superconductor
  phases are non-Abelian topological superconductors especially when
  the system is in proximity to an $s$-wave superconductor.

\item {\sl Thin films of topological Weyl semimetals.} In
  Ref.~\cite{Weyl}, it is found that in the thin film of topological
  Weyl semimetal HgCr$_2$Se$_4$ the Chern number depends on the
  thickness of the film. There is a quadratic band crossing point at
  ${\bf k}=0$ ($\Gamma$ point) when the thickness is equal to the
  critical value. Such a TBCP with $N_w=2$ is a consequence of the
  topological phase transition from normal insulator to a quantum
  anomalous Hall insulator with Chern number 2 in the system. Around
  the critical thickness the quadratic band crossing is gapped. The
  low energy Hamiltonian is given by Eq.~(\ref{iop}). All the gapped
  superconductor states are non-Abelian ones, when there is a single
  band crosses the Fermi level.

\item {\sl Optical lattices with a single quadratic band crossing.}
  Examples are checkerboard lattices near half-filling and kagome
  lattices above $1/3$ filling (or below $2/3$ filling, depending on
  the sign of the hopping)\cite{our}. When spin-polarized ultracold
  fermions are filled into the optical lattices, all the gapped
  superconductor (or superfluid) phases are non-Abelian ones in the
  weak pairing regime $|\Delta_\nu|\ll |\mu|$.

\end{itemize}

\section{Conclusion and discussions}
In this work we studied the superconducting proximity effect on
topological semimetals and related systems in the aim of searching for
Majorana fermions and non-Abelian statistics. The non-Abelian
superconductors are characterized in the bulk by an {\em odd} Chern
number which, according to bulk-edge correspondence, guarantees the
existence of one Majorana fermion in each quantized vortex. By
studying the superconducting proximity effects under general
situations, we find that for two cases A and B where a single TBCP
carries an integer winding number, {\em all} the superconducting
pairing states are non-Abelian ones. We further generalize this
property to systems: (i) where such a TBCP is gapped due to
time-reversal symmetry breaking but inversion symmetric perturbations;
(ii) when the TBCP are gapped and deformed given that only one band
crossing the Fermi level; (iii) when there are multiple such TBCPs
with an odd number of bands crossing the Fermi level if the
superconducting pairing interaction is within each TBCP; (iv) when the
TBCP system breaks time-reversal symmetry yet has inversion
symmetry. As a consequence of those findings we give several candidate
physical systems which can support the Majorana fermions that have not
attracted the attention in the community.

It is noted from Eq.~(\ref{nc}) that the Chern number changes sign
when the chemical potential moves across the TBCP, which indicates
that there is a topological phase transition in the strong pairing
regime. For superconductor and superfluid phases emerge due to
continuous phase transition driven by attractive interaction, the
gapped pairing states usually reduce the Ginzburg-Landau free energy
more than the nodal ones\cite{Anderson}. Hence the special property
found in this work may also imply that the non-Abelian pairing states
are energetically favored as the the spontaneous symmetry broken
phases. This is indeed confirmed in a subsequent work\cite{our}.

\section*{Acknowledgments}
We thank Zohar Ringel, Yuval Oreg, and the (anonymous) referee for
helpful discussions and comments.

\appendix

\section{Details of the derivation of the Chern number}

Consider, e.g., systems with a single TBCP carrying an even winding
number when $\mu>0$. There are two occupied bands of the BdG
Hamiltonian: one from the band crossing the Fermi level, and the other
one from the band below the Fermi level. In the weak pairing regime,
$|\Delta_\nu|\ll |\mu|$, one can ignore pairing between states
separated far away. One can then obtain the wavefunctions of the two
bands under the approximation
\bea
&& \Psi_{o} = e^{i\phi_{\bf k}/2}\left(\begin{array}{c}
    \sin \frac{\xi_{\bf k}}{2} \cos\frac{\phi_{\bf k}}{2} e^{i\theta_{\Delta}} \\
    \sin \frac{\xi_{\bf k}}{2} \sin\frac{\phi_{\bf k}}{2} e^{i\theta_{\Delta}} \\
    -\cos \frac{\xi_{\bf k}}{2} \cos\frac{\phi_{\bf k}}{2}  \\
    -\cos \frac{\xi_{\bf k}}{2} \sin\frac{\phi_{\bf k}}{2}
  \end{array}\right) , \nn\\
&& \Psi_{v}({\bf k}) = e^{-i\phi_{\bf k}/2}
\left(\begin{array}{c}
    \sin\frac{\phi_{\bf k}}{2} \\
    -\cos\frac{\phi_{\bf k}}{2} \\
    0 \\
    0
\label{psiov}
\end{array}\right) .
\eea
Here $\xi_{\bf k} = {\rm Arg}[\vep_{{\bf k}+} - \mu + i
|\Delta_{\rm eff}({\bf k})|]$ and $\theta_{\Delta}={\rm
  Arg}[e^{-i\phi_{\bf k}}\Delta_{\rm eff}({\bf k})]$. An important
property is that the Chern number $N_C$ does not change without
closing the gap. One can then simplify the calculation of $N_C$ by
adiabatically tuning the system. It is noted that the gap is
determined by $|\Delta_{\rm eff}({\bf k})|$ at the Fermi surface. One
can then adiabatically tune the system so that $|\Delta_{\rm eff}({\bf
  k})|$ is nonzero only in the vicinity of the Fermi
surface\cite{QiZ2}. The angular dependence $|\Delta_{\rm
  eff}(k,\theta_{\bf k})|$ (here $k_x=k\cos\theta_{\bf k}$ and
$k_y=k\sin\theta_{\bf k}$) at each energy contour can also be
adiabatically tuned to be identical to that on the Fermi surface.
The Chern number is the integration of the Berry-curvature in the
first Brillouin zone,
\be
N_C = \sum_{n=o,v} \frac{1}{2\pi}\int d{\bf k}~ {\bf e}_z\cdot
[\grad_{\bf  k}\times \langle\Psi_{n}|i\grad_{\bf k}|\Psi_{n}\rangle] .
\ee
One can divide the contribution of the integration into two parts: one
from integration over small $k$ (with a cut-off $\Lambda$), another
from integration over large $k$ region. Since there is no band-gap
closing in the large $k$ region, the Chern number is determined in the
small $k$ region, especially, in the vicinity of the Fermi surface
where the pairing gap evolves, as shown in
Ref.~\cite{QiZ2}. Inserting Eq.(\ref{psiov}), one can show that
the Chern number due to the $v$ band is zero. $N_C$ is then solely
determined by the $o$ band. Direct calculation yields
\bea
N_C &=& - \frac{1}{2\pi}\int_0^{2\pi}d{\theta_{\bf k}}\int_0^{\Lambda}d{k}
\partial_{k}\left[\sin^2\frac{\xi_{\bf k}}{2}(\partial_{\theta_{\bf k}}\theta_{\Delta})\right]
,\nn \\
&=&\left. \frac{1}{2\pi}\int_0^{2\pi}d{\theta_{\bf k}}
\partial_{\theta_{\bf k}}\theta_{\Delta}\right|_{\rm FS} ,
\eea
which results in Eq.~(\ref{nc}). Other expressions for the
Chern number $N_C$ in the main text can be derived similarly.

\end{document}